# Electronic origin of stability of 2D 1H-phase Janus transition metal dichalcogenides and beyond

Lei Li[1], Ji-Chun Lian[2,1], Zi-Xuan Yang[1], Tao Huang[1], Jun-Qi Xu[3], Jianhang Nie[1], Hui Wan[4,1], X. S. Wang[1], Gui-Fang Huang[1]†, Wangyu Hu[5], Wei-Qing Huang[1]∗

[1]*Department of Applied Physics, School of Physics and Electronics, Hunan University, Changsha 410082, China*
[2]*School of Physics and Electronic Information, Gannan Normal University, Ganzhou 341000, China*
[3]*National Laboratory of Solid State Microstructures & School of Physics, Nanjing University, Nanjing 210093, China*
[4]*School of Materials and Environmental Engineering, Changsha University, Changsha 410082, China*
[5]*School of Materials Science and Engineering, Hunan University, Changsha 410082, China*

**Abstract**: Janus transition metal dichalcogenides (JTMDs) monolayers have emerged as a new paradigm to broaden the family of two-dimensional (2D) materials. Despite numerous theoretical predictions of JTMDs, their experimental realization remains scarce, most probably due to intrinsic structural fragility. We identify a dependence of the structural stability of 1H-phase JTMDs on the transition metal group, with Group-VIB-based monolayers exhibiting robust stability, as evidenced by the successful synthesized MoSSe and WSSe. The group-dependent stability arises from the competition between metal-ligand ionic bonding and ligand-ligand covalent bonding, as well as the high-energy *d*-electron orbital splitting. We propose an electron configuration that describes the interactions of electrons near the Fermi level to correlate the stability, and introduce an electron compensation strategy to stabilize certain unstable JTMDs systems. Guided by the electronic origin of stability, we predict a family of stable 2D Janus transition metal halides with intrinsic ferromagnetic valley properties. This work bridges the gap between electronic structure and stability predictions, and extends the design rules for synthesizing 2D Janus materials.

∗. Corresponding authors: wqhuang@hnu.edu.cn
†. Corresponding authors: gfhuang@hnu.edu.cn

Two-dimensional (2D) Janus transition metal dichalcogenides (JTMDs) have attracted considerable attention since the groundbreaking synthesis of the Janus MoSSe monolayer in 2017 (*1-4*). The broken out-of-plane structural symmetry endows JTMDs with novel properties that distinguish them from their symmetrical counterparts, such as large Rashba spin-orbit coupling (*5-7*) and the emergence of strongly correlated electronic states (*8-12*). Despite numerous JTMDs and other Janus monolayers are predicted theoretically, experimental realizations remain scarce, with only two Janus monolayers—MoSSe and WSSe—successfully synthesized to date (*1, 2, 13, 14*). The origin of the challenges in the fabrication of JTMDs monolayers, in contrast to their symmetric counterparts, remains elusive (*15*).

At the atomic scale, phonon spectrum is frequently utilized to predict material stability. The presence of imaginary frequencies (soft modes) in the phonon spectrum signals structural instability. Electronic states can influence bond strengths, which in turn affect the dynamic equilibrium of atomic structure (*16-18*). For instance, Peierls distortion demonstrates how unstable phonon modes can induce structural phase transitions to a stable charge density wave (CDW) phase, thereby linking atomic vibrations with electronic structures through electro-acoustic coupling (*19-23*). Similarly, Mott transition occurs as electrons seek a lower energy configuration, ultimately leading to the formation of Mott insulators (*24-27*). Essentially, they are all electronic state rearrangements that satisfy the principle of minimum energy. Additionally, aside from thermal properties related to lattice vibrations, most material properties are governed by electrons (*28*). This highlights the disconnect between lattice dynamics-based stability predictions and electronic structure-based property predictions. Understanding the relationship between electronic structure and stability not only bridges this gap but also opens up the possibility of designing materials based on the desired electronic properties. In this work, we elucidate the origin of the stability of 1H-phase JTMDs at the electronic scale, thereby explaining the challenges in their experimental preparation. Moreover, a direct correlation between electronic properties and the prediction of structural stability is established.

**TM-group-dependent stability of 1H-phase JTMDs**. We begin by evaluating the structural stability of 84 1H-phase JTMDs with lattice constants ranging from 2.83 to 3.85 Å (Table S1), which covers nearly the entire block of transition metal (TM), comprising three periods and eight groups (14 TM atoms), in combination with four chalcogen elements (O, S, Se, and Te), as depicted in Fig. 1. Structural stability is comprehensively assessed through cohesive and formation energies for energetic stability, elastic coefficients for mechanical stability (Table S1), and phonon dispersion for dynamical stability (Fig. S1). Interestingly, the dynamically stable structures are also found to be energetically and mechanically stable. Fig. 1 summarizes the results of all calculated structures, identifying a total of 45 stable monolayers. The stability exhibits a clear dependence on the TM group: 32 stable monolayers are derived from IVB- and VIB-group TMs, 7 from IIIB- and VB-group TMs, and only 6 from VIIB- and VIIIB-group TMs.

**Bonding Characteristics in 1H-phase JTMDs**. The 1H-phase JTMDs monolayer adopts a structure where TM atoms are sandwiched between two distinct planes of chalcogen atoms, $X$ and $Y$, creating a trigonal prismatic coordination environment with $C_{3v}$ point group symmetry (Fig. 2(a)). Despite the mirror symmetry being disrupted by different ligands, the degeneracy of the wavefunction components along the $x$- and $y$-axes of the central TM persists, enabling the $d$ orbitals to split into three subgroups (Fig. 2(b)). Note that the difference in electronegativity of $X$ and $Y$ atoms weakens the $X$-$Y$ covalent bond. Even if the $X$-$Y$ covalent bond forms, its wavefunction will partially overlaps with the TM atom to establish a metal-ligand ($M$-$L$) bond, owing to the strong ionic character of TM. As shown in Fig. S2, the bonding and antibonding states of all S-Se bonds are each split into at least two distinct parts. Among these, the lower-energy bonding and antibonding states of the S-Se bonds align with the energy range of the bonding states of the $M$-$L$ bonds. This indicates that the wavefunctions of the S-Se bonds fully participate in the bonding interactions with the metal. Consequently, the ligand field orbitals of JTMDs can focus on the interactions between the $s$ and $d$ orbitals of the metal and the $p$ orbitals of ligand, forming bonding and antibonding states, as depicted in Fig. 2(c). Notably, when the $M$-$L$ bond exhibits strong ionic character, the antibonding molecular orbitals aligned with the crystal field arrangement are predominantly derived from the $d$ orbitals of the metal. This establishes a direct connection between the $d$-electron states and molecular orbital theory.

**Correlation electronic configuration and stability**. The electronic configuration of 3$d$-JTMDs monolayers is dictated by ligand field theory, as illustrated in Fig. 2(d), which reveals an increase in the number of antibonding electrons from 0 in Ti$XY$ to 6 in Ni$XY$, with Sc$XY$ exhibiting a deficiency of one bonding electron. The crystal orbital Hamilton population (-COHP) (Fig. S2) of $M$SSe monolayers corroborates these occupancy patterns. A deficiency in bonding electrons or an increase in antibonding electrons correlates with enhanced bond energy, as shown in Fig. 2(e). The magnitude of bond energy is generally a critical parameter governing material stability. However, this is invalid for the group-dependent stability of JTMDs. Instead, we identify electron interactions as the primary determinant of JTMDs stability. By analyzing the electron count difference between occupied antibonding states (OAS) and unoccupied bonding states (UBS), we categorize the monolayers from each group into the $N^n$ ($n$ = -1~6, corresponding to groups IIIB through VIIIB) system, as illustrated in Figs. 1 and 3.

A compelling correlation between magnetic moments and stability is identified in $N^{-1}$ to $N^2$ systems. Most $N^0$ and $N^2$ monolayers are stable with zero magnetic moments, while some $N^{-1}$ and $N^1$ structures achieve stability with magnetic moments of 1 μB (Fig. 1). In contrast, structures exhibiting fractional magnetic moments are generally unstable, indicating that integer magnetic moments could enhance stability by allowing certain structures to remain stable despite a higher number of antibonding electrons. In fact, the integer magnetic moment indicates orbital splitting around the Fermi surface. In particular, the complete splitting of energy levels at the Fermi level ($E_F$) induces semiconducting behavior, thereby strengthening the structural stability of the system.

For the $N^{-1\sim2}$ systems, the low-energy bonding and antibonding states of the $X$-$Y$ bond are almost fully occupied, resulting in a near-zero bond order (except for oxygen-free Sc$XY$, which will be discussed later). The $M$-$L$ bond is primarily filled with bonding states, leading to a large bond order, as depicted in Fig. 3. In this case, the stability of these structures is primarily governed by the $M$-$L$ ionic bond, with antibonding states predominantly composed of $d$ orbitals. These orbitals are particularly susceptible to repulsive interactions, thus promoting energy level splitting. We identify three stable configurations, $N^0$, $N^1$, and $N^2$, each associated with a distinct splitting mechanism: the molecular orbital splitting (MOS), spin polarization splitting (SPS), and crystal field splitting (CFS), respectively.

However, for the $N^{3\sim6}$ systems, fractional magnetic moments dominate, and most structures exhibit intrinsic instability. The weak ionic nature of late TM introduces $X$-$Y$ covalent bond components decoupled from the central TM within the ligand field, while simultaneously diminishing the $d$-orbital contribution to the antibonding states in $M$-$L$ bonds. This leads to irregular magnetic moments and suppressed splitting near the $E_F$. Moreover, the progressive filling of antibonding states with increasing electron count reduces the ionic bond order (Fig. 3), further destabilizing the structures and driving their intrinsic instability.

Conversely, the stability of certain structures improves as the high-energy $X$-$Y$ covalent bonding states become increasingly occupied, raising their bond order. Notably, monolayers such as NiSSe, NiSeTe, and NiSTe exhibit the highest bond orders due to their fully populated $X$-$Y$ bonding states, which even surpass the ionic bond contributions, thereby conferring structural stability. However, the strength of $X$-$Y$ bond is influenced by the electronegativity difference between the ligands. For instance, systems containing oxygen are uniformly unstable because of the substantial electronegativity discrepancy between oxygen and other ligands. The competition between reduced ionic bond order and strengthened covalent interactions plays a crucial role in stability. In compounds like MnSeTe, FeSSe, and FeSTe, the relatively high covalent bond order compensates for the diminished ionic bond order, allowing these systems to retain stability.

**MOS for stability of $N^0$ system**. The orbital splitting of $N^0$ configurations is depicted in Fig. 4(a), alongside the $N^2$ configuration for comparison. In the $N^0$ configuration, the ligand $p$ orbitals are fully occupied, while the metal $d$ orbitals remain entirely unoccupied, exemplified by TiOS (Fig. 4(b)). This results in a clear separation between bonding and antibonding states at the $E_F$, yielding stable structures with minimized bond energy. Band structures (Fig. S3) and -COHP analyses (Fig. S4) reveal that most structures are stable in the $N^0$ systems, with exceptions such as ZrOTe and HfOTe, where incomplete splitting at $E_F$ results in fractional magnetic moments and thus instability. Interestingly, 1H-ZrSSe is unstable despite complete MOS, likely due to phase competition. As shown in Fig. S5, the integrated -COHP of 1T-ZrSSe below $E_F$ is lower than that of 1H-ZrSSe, favoring a phase transition to the energetically more stable structure.

In fact, the degree of MOS is governed by the energy alignment between the lower edge of the metal's $d$ band and the upper edge of the ligand's $p$ band, as shown in Fig. 4(c1). For instance, in TiOS, the CBM primarily originates from the metal's $d$ orbitals, while the VBM arises from the ligand's $p$ orbitals (Fig. S6(a)). Typically, $d$ orbitals are localized, whereas $p$ orbitals are more diffuse. Excessive broadening of $p$ orbitals can result in crossing the $E_F$ and forming metallic bands, as depicted in Fig. 4(c2). This phenomenon is exemplified by Hf$XY$ systems. In HfO$Y$, the high electronegativity of oxygen reduces $p$ orbital overlap compared to other ligands (Fig. S6(b1-b2)). Additionally, from S to Se to Te, $p$ orbital broadening increases, with Te $p$ orbitals crossing the $E_F$. Consequently, HfOTe displays metallic characteristics (Fig. S6(c)), where partially occupied Hf-$d$ orbitals and unoccupied ligand-$p$ orbitals lead to incomplete filling of bonding and antibonding states (Fig. S6(d)), thereby compromising stability. Therefore, the complete MOS at the $E_F$ is necessary for stability of $N^0$ system.

**CFS for stability of $N^2$ system.** The $N^2$ system possesses two additional antibonding electrons compared to the $N^0$ system, ostensibly suggesting reduced stability. Paradoxically, the monolayers in $N^2$ system are stable except for CrOTe (Fig. 1), due to a distinct splitting mechanism. As illustrated in Fig. 4(a), for the $N^2$ configuration, the $a^*$ and $e_1^*$ orbitals are driven apart at the $E_F$ by CFS. This splitting is predominantly governed by the central TM atom and is largely independent of the ligand, as evidenced by the CrOS monolayer (Fig. 4(d)). Moreover, the Jahn-Teller (JT) effect plays a pivotal role in lowering the energy of the $a^*$ orbital through bond length modulation, thus strengthening orbital splitting. As shown in Fig. S7-S8, the significant energy reduction of the $a^*$ orbital in $N^2$ systems reduces the MOS between the $a^*$ and ligand $p$ orbitals, whereas the unoccupied $e_1^*$ and $e_2^*$ orbitals remain energetically unchanged. Consequently, the JT effect further amplifies the energy difference between the $a^*$ and $e_1^*$ orbitals, reinforcing the CFS. The combined effect of CFS and the JT effect leads to an upward shift of high-energy antibonding states and a downward shift of low-energy antibonding and bonding states, thereby lowering the total energy and enhancing the structural stability of the $N^2$ system. However, as a special case within the $N^2$ system, monolayer CrOTe is unstable due to the absence of a bandgap induced by CFS (Fig. S7).

The magnitude of the formed bandgap can serve as an indicator of the degree of splitting and substantiate the correlation between the stability and splitting of the $N^2$ system. As depicted in Fig. 4(e), from Cr$XY$ to Mo$XY$ to W$XY$, the bandgap progressively increases as the central TM atom becomes heavier, while bond energy decreases, highlighting the critical role of pronounced CFS in structural stability. Notably, Mo$XY$ and W$XY$ exhibit superior structural stability compared to Cr$XY$, aligning with experimental results that only Janus MoSSe and WSSe monolayers have been successfully synthesized to date. This addresses longstanding challenges in experimental synthesis of other JTMDs monolayers.

**SPS for stability of $N^1$ system and its predictive application.** Despite possessing only one antibonding electron, most monolayers of $N^1$ system are inherently unstable, characterized by fractional magnetic moments and the absence of a bandgap, as depicted in Fig. 5(a). However, four monolayers—VSSe, VSeTe, VSTe, and VOSe (Fig. S9) — exhibit a stable $N^1$ configuration, each demonstrating a magnetic moment of 1 μB. Taking VOSe as an example (Fig. 5(b)), complete spin polarization splits the $a^*$ orbital into spin-up and spin-down states at the $E_F$. This process induces

CFS in the spin-up band and MOS in the spin-down band, as illustrated in Fig. 5(c1). Consequently, the stability of $N^1$ configuration monolayers depends on the synergy of these two distinct splitting mechanisms. Specifically, since the CFS energy of the $N^1$ system lies intermediate between those of the $N^0$ and $N^2$ systems, the spin-down $a^*$ orbital would be pushed down compared to $N^0$ systems, making MOS more susceptible to being eliminated. Simultaneously, the energy gap between the spin-up $a^*$ orbital and the high-energy antibonding orbitals is reduced compared to $N^2$ systems, making the CFS more likely to be eliminated, as shown in Fig. 5(c2). Therefore, the weakening of both mechanisms leads to the intrinsic instability of most monolayer structures in the $N^1$ system (Fig. 1).

Interestingly, the instability of the $N^1$ system can trigger acoustoelectric coupling, driving a structural transition into a stable charge density wave (CDW) phase (Fig. 5(d)) (*29-33*). Such a phase transition is well-documented in many 1H-phase TMDCs, which frequently undergo transformations into CDW structures. This phase transition process highlights the governing role of the minimum energy principle at the electronic scale in facilitating atomic-scale structural stabilization. Strikingly, the resulting CDW phase adopts the $N^1$ configuration, aligning with the earlier analysis of the $N^1$ system's stability.

The degree of SPS critically determines the feasibility of forming the $N^1$ configuration. Any energy overlap between two spin components of the $a^*$ orbital eliminates orbital splitting at the $E_F$. However, the suppression of the spin polarization field by the crystal field, combined with the broader energy distribution of the $a^*$ orbital, makes achieving complete SPS at the $E_F$ challenging, particularly in Nb- and Ta-based $N^1$ systems.

The $N^1$ configuration offers a conceptual framework for designing novel, stable monolayer structures. Substituting Group-VIA ligands with halogen ligands and employing Group-IIIB metals as central atoms yields 18 stable Janus transition metal dihalogen compounds (JTMDHs): Sc (Y, La) *X'Y'* (*X'*, *Y'*=F, Cl, Br, I, *X'*≠*Y'*) monolayers (Fig. S9). In these JTMDHs, the crystal field effect is weakened, while SPS at the $E_F$ is enhanced. Furthermore, the increased ionic character of *M–L* bonds strengthens *d* orbital repulsion in antibonding states. Consequently, the Group-IIIB JTMDHs monolayers adopt $N^1$ configurations (Fig. S9), with calculated bandgaps (Fig. 5(a)) and phonon dispersion results (Fig. S10) affirming their structural stability. These stable Group-IIIB JTMDHs monolayers are ferromagnetism semiconductor with a magnetic moment of 1 μB, presenting a promising avenue for diverse applications (*34*). For instance, Sc-based monolayers exhibit naturally ferromagnetic valley polarization (Fig. S11) due to intrinsically centrosymmetry breaking (*35-37*). In spin-up bands of ScBrI, the degeneracy at the *K* and *K'* valleys of VBM is lifted, yielding an energy difference of 69 meV (Fig. 5(f1)) and demonstrating valley polarization. Moreover, the Berry curvature distribution (Fig. 5(f2)) exhibits identical magnitudes with opposite signs at the *K* and *K'* valleys, validating the concept of valley-contrasting Berry curvature.

**Electron self-compensation and its application**. For the six monolayers in the $N^{-1}$ system, those without oxygen are stable, while those containing oxygen are all unstable. Fig. 6(a) shows that

ScOS exhibits metallic behavior due to an unoccupied state in the ligand's $p$ orbital, a characteristic shared by ScOSe and ScOTe (Figs. S12(a1)). In contrast, SPS causes the ligand $p_z$ orbitals in oxygen-free Sc$XY$ monolayers to distribute on both sides of the $E_F$, exhibiting semiconductor characteristics (Figs. 6(b), S12(a2)). Furthermore, PDOS analysis reveals that the polarized orbitals near the $E_F$ originate equally from the $p_z$ orbitals of S and Se (Fig. 6(b)), and the wavefunctions at CBM and VBM contributed by S and Se but almost no $d$ orbital of Sc (Fig. 6(c)), indicating the formation of $\sigma$ bonds between the ligand $p_z$ orbitals.

The large atomic radius of Sc leads to an expanded lattice constant of the Sc$XY$ system relative to other systems (Figs. S12(b)). Due to the Poisson effect, the lattice expansion induces $xy$-plane stretching and $z$-direction compression, shortening the $X$-$Y$ distance (Fig. 6(d)) (*38-40*). Consequently, pairwise $\sigma$ bonds form between S, Se, and Te. However, the significant energy differences between O and other ligands hinder O-$Y$ bond formation. Therefore, the presence of UBS renders ScO$Y$ system unstable (Fig. 6(e)). By contrast, ScSSe, ScSeTe, and ScSTe monolayers form $\sigma$ bonds that share two electrons, which is equivalent to adding two electrons, i.e., electron self-compensation. One electron fills the UBS, while the other occupies an antibonding $\sigma^*$ orbital. Remarkably, the $\sigma^*$ orbital features alternating nodes in the $z$-direction (Fig. 6(c)) and exhibits localization in the $xy$-plane, minimizing the effects of periodic modulation and producing a flat band with highly localized electrons within the Brillouin zone. SPS results in distinct spin components for CBM and VBM (Fig. 6(b)), resulting in a stable $N^1$-like configuration.

The electron self-compensation motivates the implementation of hydrogenation as an effective compensation strategy. Introducing hydrogen at the oxygen site forms H-ScO$Y$, which not only provides electron compensation but also weakens the Sc-O bond. As illustrated in Fig. 6(f), as hydrogen donates an electron to oxygen, Sc only needs to provide three electrons, thereby achieving a $N^0$ configuration. Band structure and -COHP analyses in Fig. 6(g) reveal fully occupied bonding $p$ orbitals and completely empty antibonding $d$ orbitals, resulting in the formation of a bandgap and confirming the $N^0$ configuration. Phonon spectrum calculations further validate the stability of H-ScOS, H-ScOSe and H-ScOTe monolayers (Fig. S12(c1-c3)). Notably, this hydrogenation strategy is not limited to the $N^{-1}$ system, but can be broadly applied to stabilize various monolayers within the $N^0$ to $N^1$ system. For instance, H-ZrOTe transitions into a stable $N^1$ configuration (Fig. S12(d)), while H-VOS and H-VOTe undergo transformations into a stable $N^2$ configuration (Figs. S12(e1, e2)).

In summary, we elucidate the electronic origins of structural stability in 1H-phase Janus JTMDs. The group-dependent stability originates from the competition between metal-ligand ionic bonding and ligand-ligand covalent bonding, as well as the antibonding states splitting at the Fermi level through $d$ orbitals repulsion. Three distinct electronic configurations are identified as the primary determinant of stability, corresponding to MOS, CFS, and SPS, respectively. We propose an electron compensation strategy to enable stabilization of unstable JTMDs and predict a new family of 2D Janus transition metal halides with intrinsic ferromagnetic valley properties. This work bridges the gap between electronic structure and stability predictions, offering a practical guidance for synthesizing JTMDs and beyond.

**Data Availability Statement**

The data that support the findings of this study are available from the corresponding author upon reasonable request.

**Supplementary Materials**

See the supplementary materials for the computational methods and additional results.

**Acknowledgements**

The authors are grateful to the National Natural Science Foundation of China (Grants No. 11804045, 12174093 and 52172088) and the Fundamental Research Funds for the Central Universities.

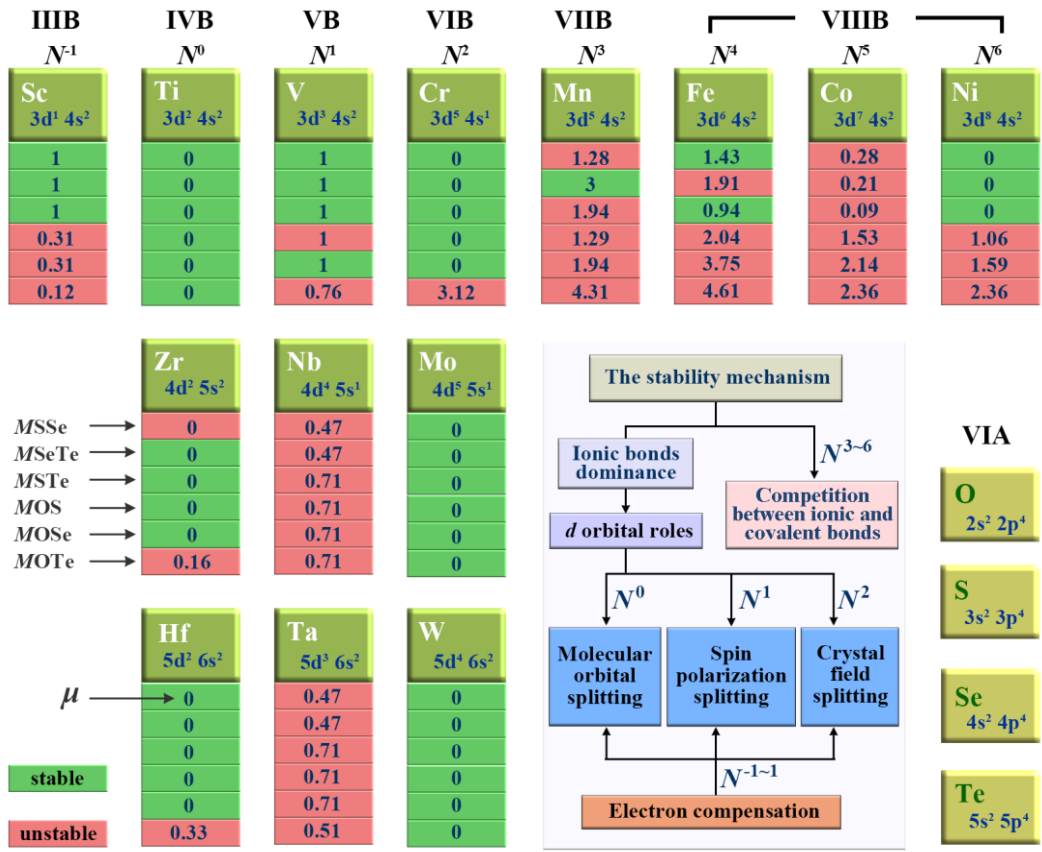

**Fig. 1. The group-dependent stability, magnetic moment, and stability mechanism diagram.** The structure stability and magnetic moment of 84 JTMDs $MXY$ ($M$=Sc, Ti, V, Cr, Mn, Fe, Co, Ni, Zr, Hf, Nb, Ta, Mo, W). The inset provides an outline diagram summarizing the stability mechanism.

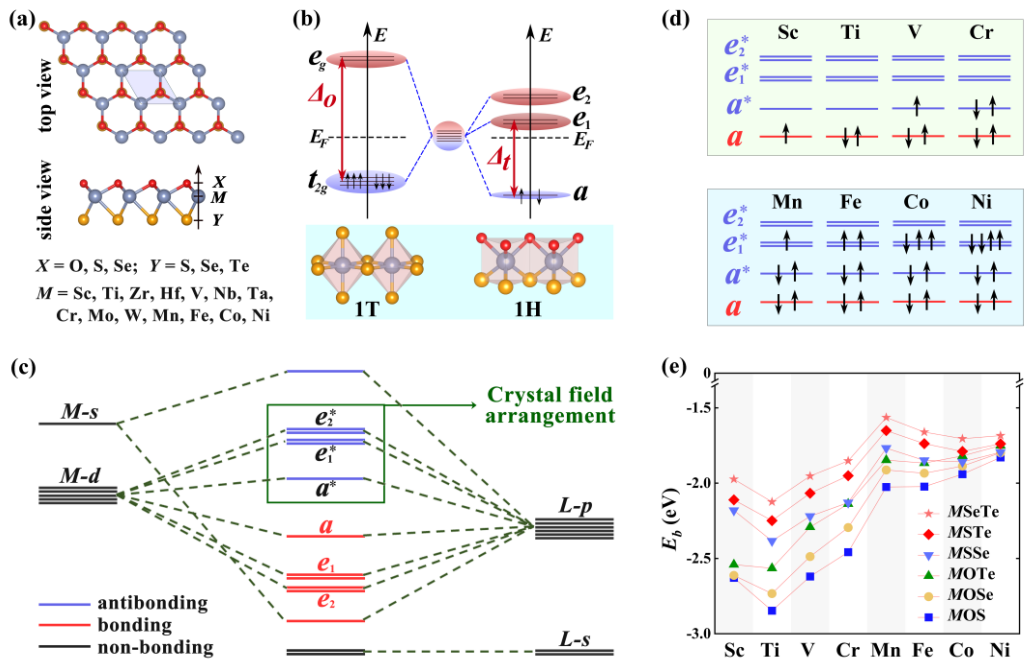

**Fig. 2. Stability analysis based on ligand field.** (a) The top and side views of 84 1H JTMDs. (b) The crystal field of 1T and 1H phase JTMDs. (c) The molecular orbital diagram in ligand field of JTMDs. (d1, d2) The filling of molecular orbitals for 3$d$-JTMDs with the corresponding center metal marked above. (e) The bond energy of 3$d$-JTMDs.

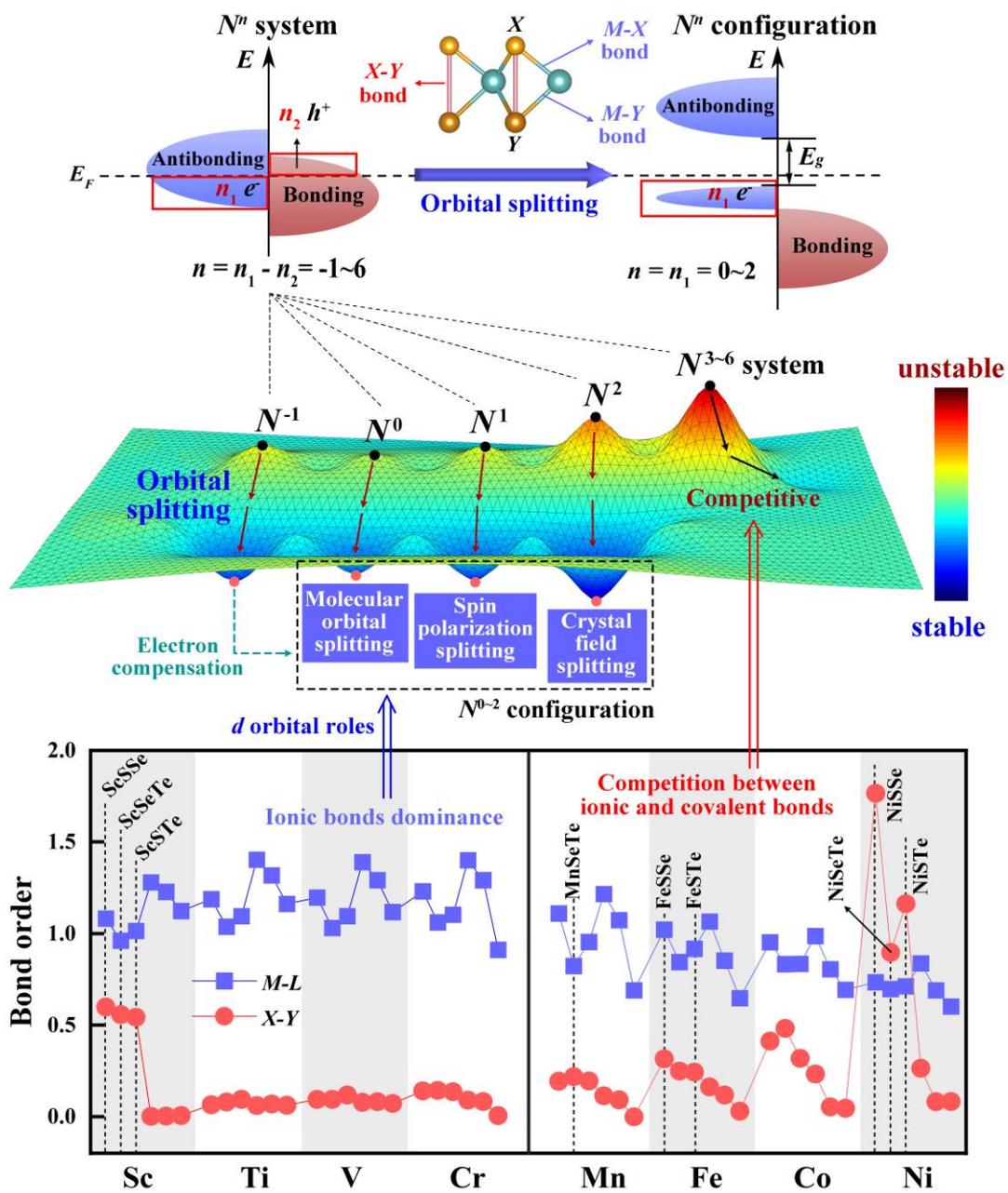

**Fig. 3. The stability mechanism diagram.** The diagram of $N^n$ system and $N^n$ configuration, the stability mechanism diagram, and the bond order for *M-L* bonds and *X-Y* bonds of JTMDs.

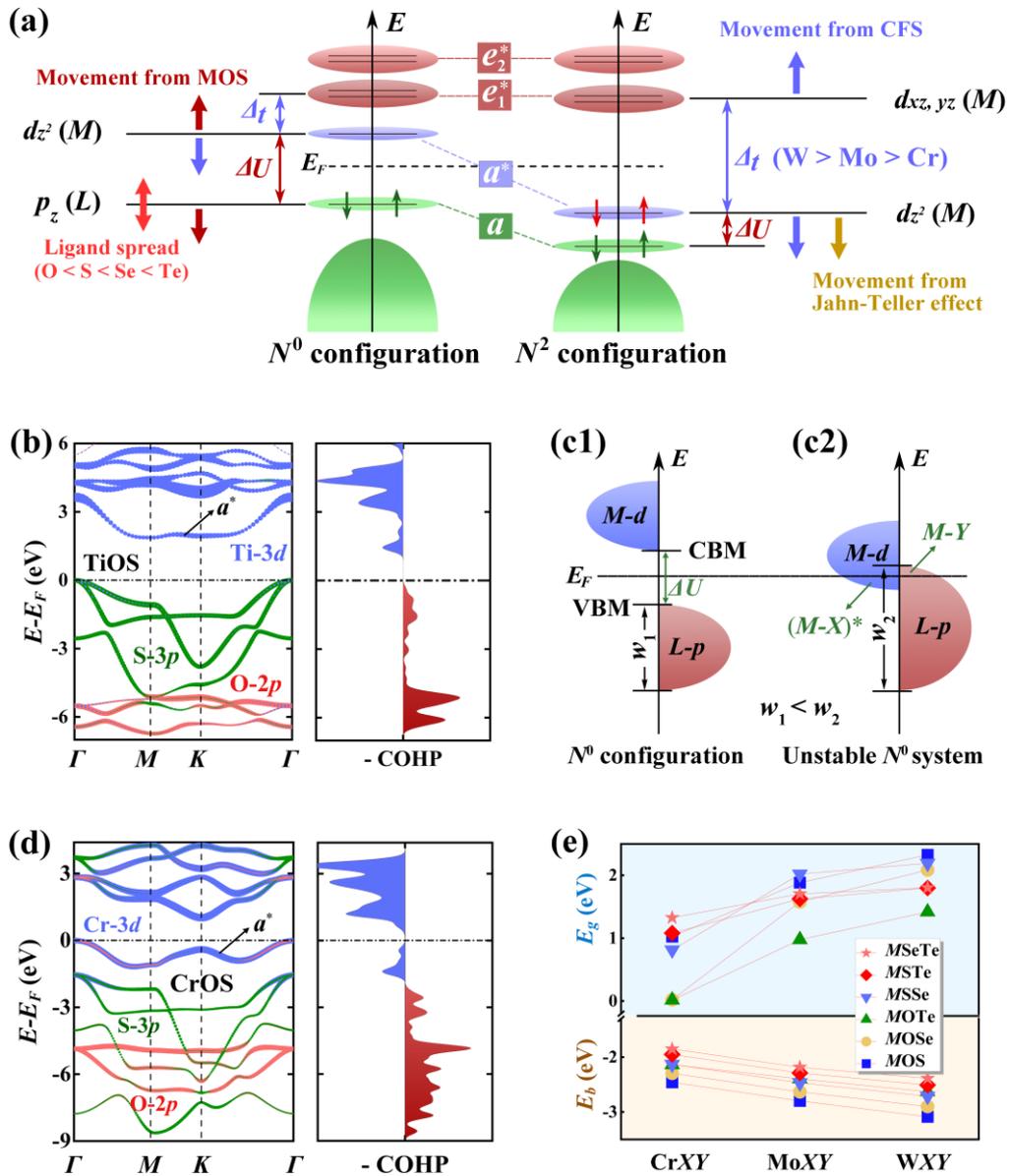

**Fig. 4. Stability based on MOS and CFS.** (a) The orbital splitting diagram of $N^0$ and $N^2$ configuration. (b) The projected energy band of TiOS and the average -COHP of Ti-O and Ti-S bonds. The molecular orbital diagram of (c1) $N^0$ configuration, and (c2) unstable $N^0$ system with metal band. (d) The projected energy band of CrOS and the average -COHP of Cr-O and Cr-S bonds. (e) The variation trend of band gap and bond energy for $N^1$ systems.

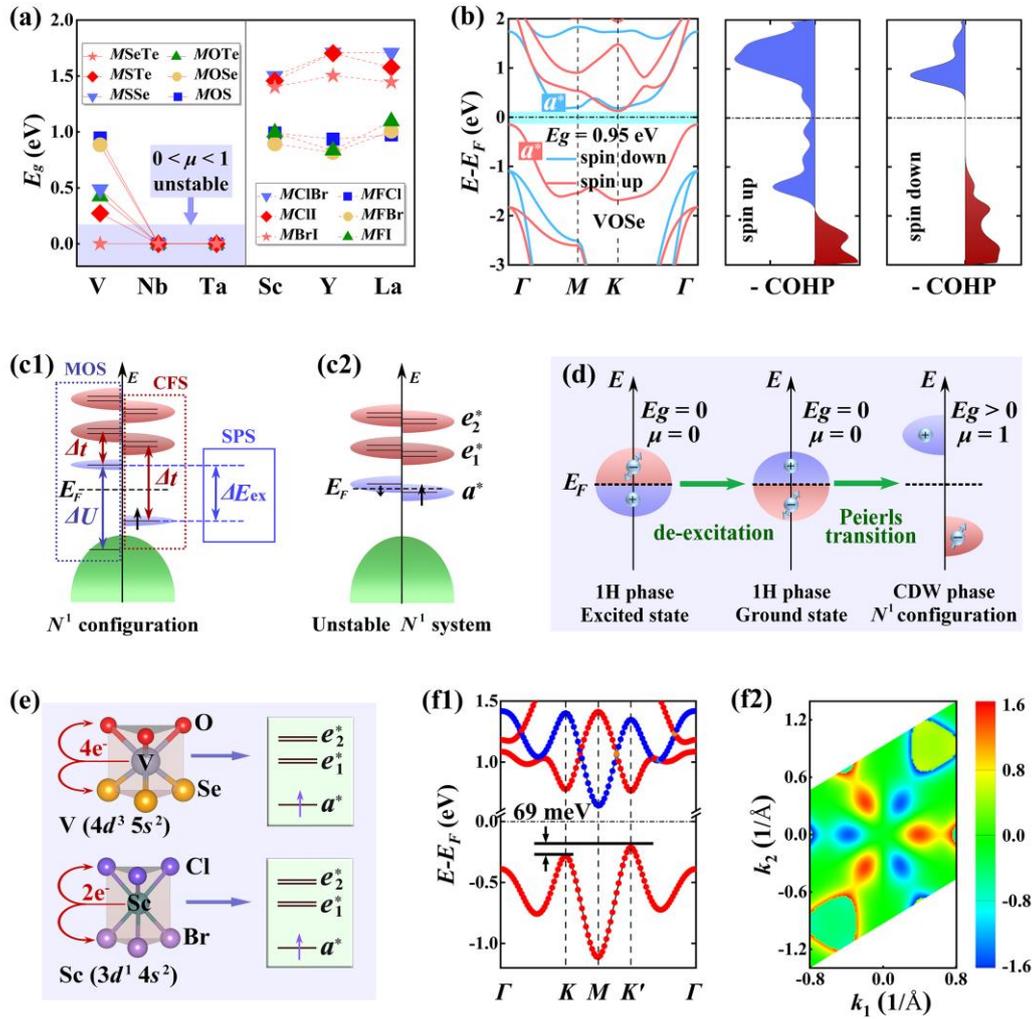

**Fig. 5. Stability based on SPS.** (a) The band gap of Group-VB JTMDs and Group-IIIB JTMDHs. (b) The projected energy band of VOSe and average -COHP of V-O and V-Se bonds. Schematic diagram of (c1) spin polarization splitting of $N^1$ configuration, and (c2) unstable $N^1$ system with metal band. (d) The schematic diagram of Peierls transition. The schematic diagram of electron configuration for (e) VOSe and ScClBr. (f1) Spin-polarized energy band of ScBrI with SOC near the $E_F$. Red and blue lines indicate spin-up and spin-down states. (f2) The contour map of the Berry curvature distributions 2D BZ for ScBrI monolayer.

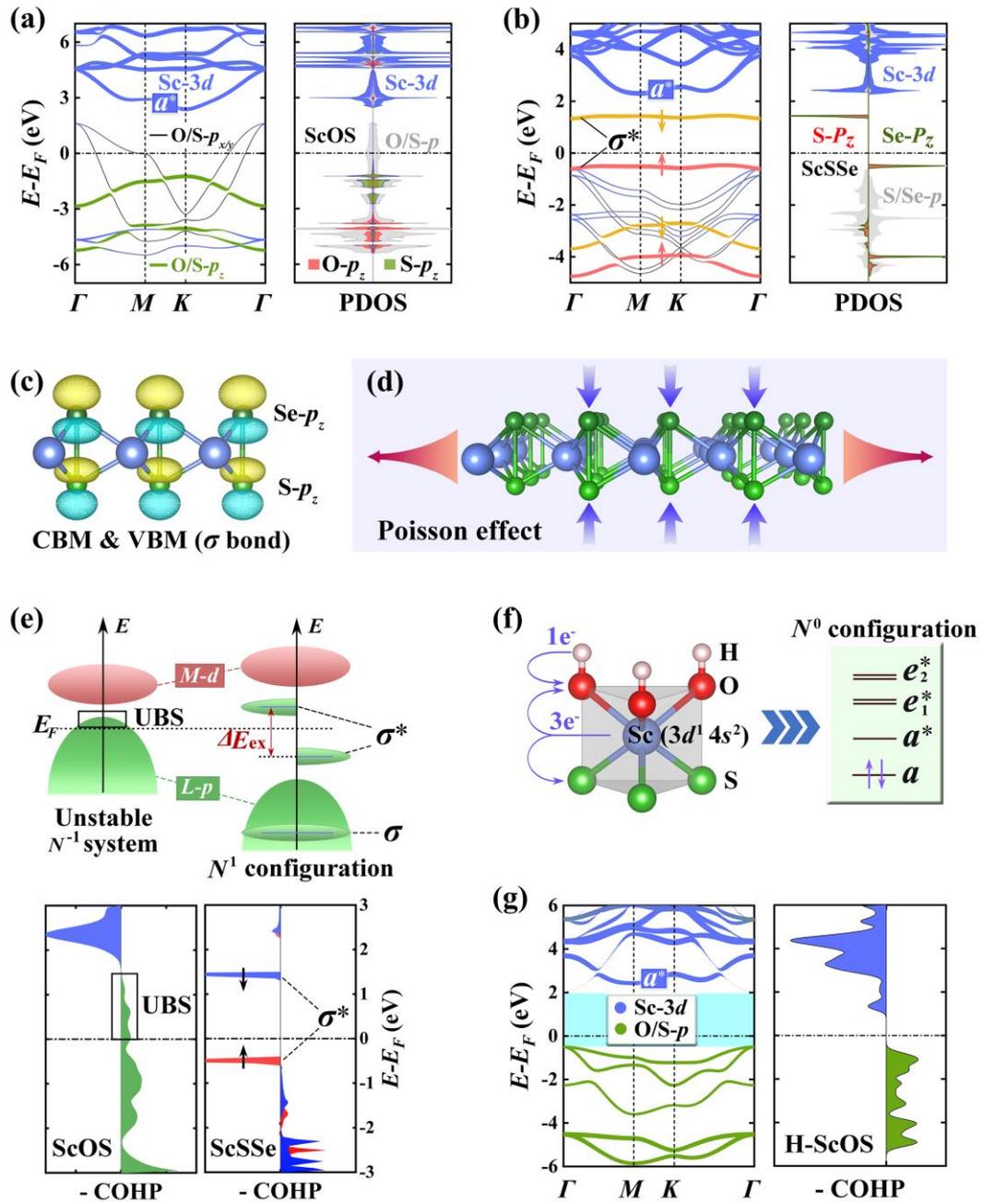

**Fig. 6. Electronic compensation effect induced stability.** The projected energy band and density of state of (a) ScOS and (b) ScSSe. (c) The same wave function in CBM and VBM of ScSSe. (d) Diagram of Poisson effect. (e) Schematic diagram of unstable $N^{-1}$ system and stable $N^1$ configuration, and corresponding average -COHP of ScOS and ScSSe. (f) The schematic diagram of electron configuration of H-ScOS. (g) The projected energy band of ScOS and average -COHP of Sc-O and Sc-S bonds.